\pdfoutput=1
\documentclass[%
 reprint,
 amsmath,amssymb,
 aps,
]{revtex4-2}

\usepackage{graphicx}
\graphicspath{ {./figures/} }
\usepackage{dcolumn}
\usepackage{bm}

\usepackage[utf8]{inputenc}
\DeclareUnicodeCharacter{2212}{-}

\begin{document}

\preprint{APS/123-QED}

\title{Stratification of polymer-colloid mixtures via fast nonequilibrium evaporation}


\author{Kyoungmun Lee}
 \email{lkm2387@kaist.ac.kr}
 \author{Siyoung Q. Choi}%
 \email{sqchoi@kaist.ac.kr}
\affiliation{%
Department of Chemical and Biomolecular Engineering, Korea Advanced Institute of Science and Technology (KAIST), Daejeon 34141, Korea
}%

\date{\today}

\begin{abstract}
In drying liquid films of polymer-colloid mixtures, the stratification in which polymers are placed on top of larger colloids is studied. It is often presumed that the formation of segregated polymer-colloid layers is solely due to the proportion in size at fast evaporation as in binary colloid mixtures. By comparing experiments with a theoretical model, we found that the transition in viscosity near the drying interface was another important parameter for controlling the formation of stratified layers in polymer-colloid mixtures. At high evaporation rates, increased polymer concentrations near the surface lead to a phase transition from semidilute to concentrated regime, in which colloidal particles are kinetically arrested. Stratification only occurs if the formation of a stratified layer precedes the evolution to the concentrated regime near the drying interfaces. Otherwise, the colloids will be trapped by the polymers in the concentrated regime before forming a segregated layer. Also, no stratification is observed if the initial polymer concentration is too low to form a sufficiently high polymer concentration gradient within a short period of time. Our findings are relevant for developing solution-cast polymer composite for painting, antifouling and antireflective coatings.

\end{abstract}

\maketitle

\section{\label{sec:level1}Introduction}
Solution-cast polymer composite films composed of polymer matrices containing colloidal particles have been widely studied for many applications, including paints [1], coatings [2,3], and cosmetics [4,5] because they provide highly improved macroscopic properties relative to the pure polymer [6], through a simple manufacturing process. The enhanced properties of the dried films are largely dependent on the spatial distribution of the polymer and colloid [7-10]. In particular, stratified layers consisting of a polymer layer on a colloidal layer have exhibited highly improved antifouling performance [11,12], and photoactive properties [13].

Several previous studies have demonstrated ways of controlling the segregated layers of polymer-colloid mixtures in an equilibrium state [14-16]. However, relatively little is known about how polymer-colloid mixtures can be stratified during the simple, fast and inexpensive nonequilibrium solvent evaporation process. Although solvent casting is one of the simplest manufacturing methods, from coffee ring stains [17] to many industrial applications [1-5], the inherent nonequilibrium nature of drying has made it difficult to clarify the underlying mechanism.

As a solvent evaporates, the spatial distribution of the solutes in liquid films is determined by two competing factors: diffusion [18] and receding drying interfaces. Solutes tend to distribute uniformly in drying films with a diffusion constant \emph{D}, while the nonuniform concentration gradient is developed by the downward velocity of the interface \emph{$v_{ev}$}. Which of the two phenomena dominates can be quantified by the dimensionless \emph{P\'eclet} number \emph{Pe =} \emph{$v_{ev}$}\emph{$z_0/D$}, where \emph{$z_0$} is the initial film thickness. If \emph{Pe} $>$ 1, the solutes cannot diffuse uniformly within the time of evaporation, and they accumulate near the top of the film. On the other hand, the drying film shows almost uniform distribution if \emph{Pe} $<$ 1.

In binary colloid mixtures, it was recently shown that stratifications with smaller colloids placed on large colloids can be realized if \emph{Pe} is larger than 1 [19-22]. This occurs when the concentration gradient of both the large and smaller particles increases near the liquid/air interface. Fortini \emph{et al.} [20] proposed that the inverted stratification was caused by an imbalance in the osmotic pressure between the larger and smaller colloids. Zhou \emph{et al.} [21] suggested that the stratification phenomenon could be explained quantitatively using a diffusion model, with cross-interaction between the colloids. Sear and Warren [22] argued that diffusiophoretic motion induced by the concentration gradient of the smaller components can exclude the larger colloids from the drying interfaces.

In a way similar to binary colloid mixtures, it has been proposed that a polymer-colloid mixture can yield the same stratified layers if the \emph{Pe} of both the polymer and colloid are larger than 1 [23,24]. However, these results have only been demonstrated by simulation and modeling studies, and few experimental studies have been made on polymer-colloid stratification. Although polymers and colloids can show similar behaviors at very dilute concentrations [24,25], they might behave much differently at the high concentrations that any drying solutions must experience for the complete drying [26,27]. The obvious difference is viscosity. It rapidly increases at relatively low concentrations in the polymer solution, slowing the motions of the species [27-29]. In contrast, the viscosity of the colloidal suspension increases relatively slowly [30]. Thus, the growth in viscosity near the interface, which can kinetically arrest larger colloids [31-33], needs to be considered differently for polymer and colloidal systems, but no appropriate studies have been performed yet.

In this work, we experimentally show that the formation of stratified layers, where a small polymer layer is placed on larger colloids, can be predicted using two competing time scales: the time at which the colloid begins to stratify (\emph{$t_s^*$}) and the time the colloid is arrested by the transitions of viscosity near the interface (\emph{$t_c^*$}).

We consider that the colloid starts to be arrested near the drying interfaces when the polymer concentration reaches a concentrated regime where the polymer chains are densely packed [29]. The stratification can be observed only if \emph{$t_s^*$} precedes \emph{$t_c^*$}, or \emph{$t_c^*$}/\emph{$t_s^*$} $>$ 1. Otherwise, the viscosity near the drying interface rapidly grows within a very short time and the colloids are kinetically trapped before a sufficient downward velocity away from the surface of large colloids is generated. In addition, when the initial polymer concentration is too low, no stratification can also occur because the concentration gradient of the polymer, or the additional migration velocity of the larger colloid, is not enough until the evaporation ends.

For the predictive analysis of \emph{$t_s^*$} and \emph{$t_c^*$}, we propose a simple model modified from the previous work [22]. We observed quite excellent agreement in the final film morphology of the model prediction and experimental studies. Our comprehensive study predicts the spatial distribution of polymers and colloids in the final dried film, based on the experimental system and drying conditions.

\section{Result and discussion}

\subsection{\label{sec:level2}Structure of dried films of polymer-colloid}
Mixtures of aqueous polystyrene (PS) suspension with a mean diameter \emph{$d_c$} = 1 \emph{$\mu$}\emph{m}, and poly(ethylene glycol) (PEG) or poly(vinyl alcohol) (PVA) were used as a model system for stratification. The molecular weights of the polymers with PEG \emph{$M_n$} (number average molecular weight) 6,000 gmol$^{-1}$, PEG \emph{$M_n$} 20,000 gmol$^{-1}$, PVA \emph{MW} 6,000 gmol$^{-1}$, and PVA \emph{$M_w$} (weight average molecular weight) 13,000-23,000 gmol$^{-1}$ (PVA \emph{$M_w$} 18,000) were chosen for radius of colloid (\emph{$R_{colloid}$}) \emph{$\gg$} radius of polymer (\emph{$R_{polymer}$}). Before drying, the film solutions contained an initial volume fraction of \emph{$\phi_{i,p}$} = 0.01 or 0.04 for the polymer and \emph{$\phi_{i,c}$} = 0.67\emph{$\phi_{i,p}$} for the colloid, respectively. The mixture solutions were deposited on glass substrates as \emph{$z_0$} = 1.25 mm. The evaporation was performed at ambient temperature and a relative humidity of 23 \%, resulting in an initial polymer \emph{P\'eclet} number \emph{$Pe_{i,p}$} $>$ 1 (See Supplemental Material). All of the experimental systems are summarized in Table I. When the evaporation was completed, the final film morphologies were analyzed with the help of scanning electronic microscopy (SEM) and ImageJ analysis.

\begin{table*}
\caption{\label{tab:table3}Various polymer-colloid systems that were tested. Colloid was fixed as PS to exclude gravitational effect during drying ($\rho_{PS}$ $\approx$ $\rho_{water}$). A total of 8 systems were experimentally performed.}
\begin{ruledtabular}
\begin{tabular}{c c l c c c c c c c}
 
 &&&&&&&&
 \multicolumn{2}{c}{\emph{$Pe_{i,p}$}} \\
 \cline{9-10}
 
 Colloid & 
 Polymer &&
 \emph{$R_g$}\footnote{See Supplemental Material} (nm) & 
 \emph{$\phi_{i,p}$} & 
 \emph{$\phi_{i,p}:\phi_{i,c}$} & 
 \emph{$h_0$} (mm) &
 Relative humidity & 

 \emph{$\phi_{i,p}$} 0.01 &
 \emph{$\phi_{i,p}$} 0.04 \\ \hline
 
 PS (r = 500 nm) & PEG & \emph{$M_n$} 6,000 & 3.6 & 0.01 & 3 : 2 & 1.25 & 23 \% & 4 & 7 \\
  & & \emph{$M_n$} 20,000 & 7.4 & or & & & & 9 & 22 \\
  & PVA & \emph{MW} 6,000 & 3.5 & 0.04 & & & & 4 & 9 \\
  & & \emph{$M_w$} 18,000 & 6.8 & & & & & 8 & 24
  
\end{tabular}
\end{ruledtabular}
\end{table*}

After complete drying, the polymers were enriched at the top of the films in PEG \emph{$M_n$} 6,000 gmol$^{-1}$ (\emph{$\phi_{i,p}$} = 0.04) [Fig. 1(a)] and PVA \emph{MW} 6,000 gmol$^{-1}$ (\emph{$\phi_{i,p}$} = 0.04) [Fig. 1(c)] while other 6 dried films in Fig. 1(b), 1(d) and Fig. 2(a) - 2(d) were not segregated, but randomly distributed. Although the stratified layers in Fig. 1(a) and Fig. 1(c) also showed different degrees of stratification, there was a clear boundary between the stratified layers [Fig. 1(e)] and nonstratified layers [Fig. 1(f), Fig. 2(e), and Fig. 2(f)].

\begin{figure}
\includegraphics[scale = 0.7]{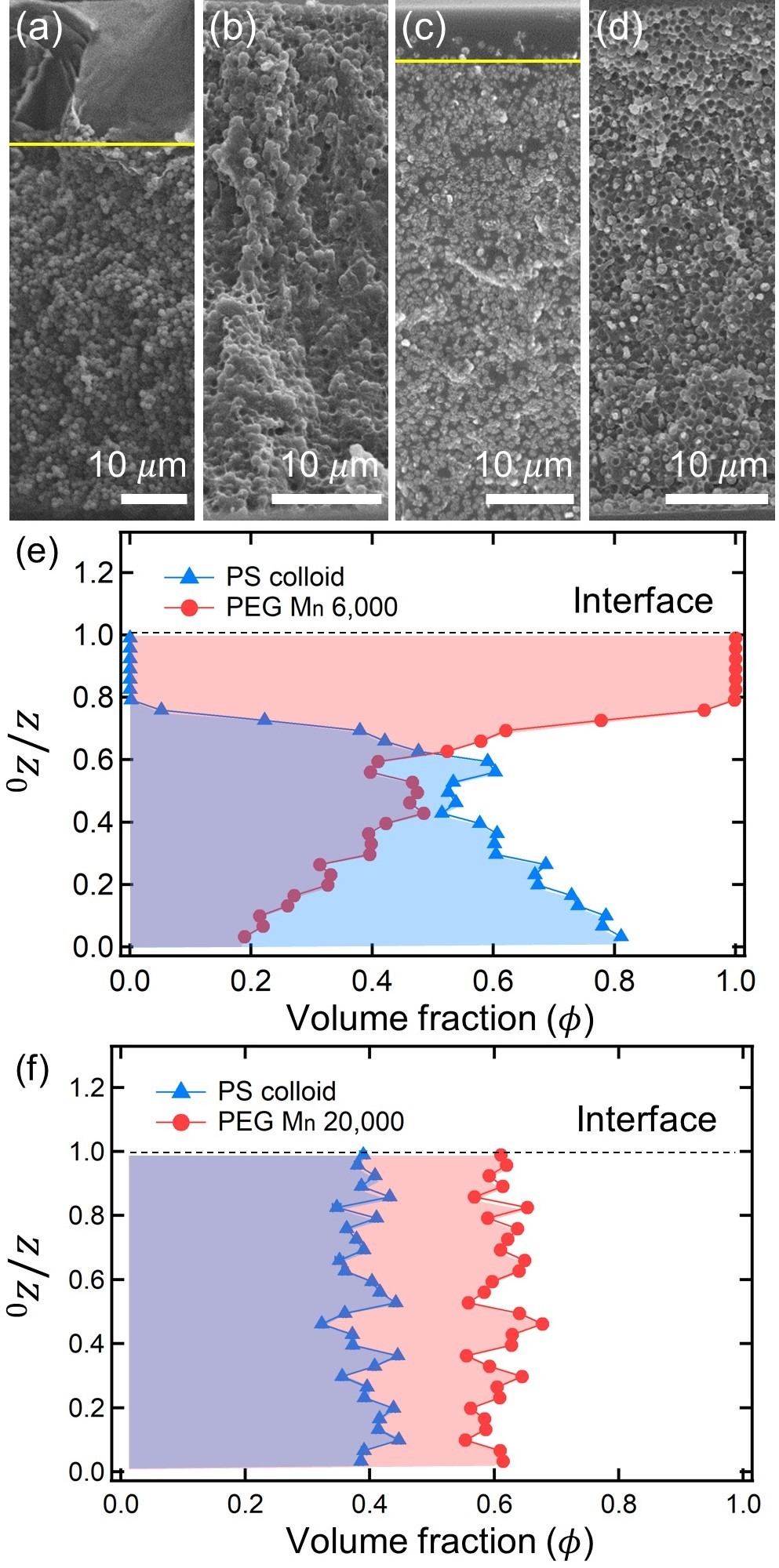}
\caption{\label{fig:epsart}Cross sectional SEM images of dried films of polymer-colloid mixtures (\emph{$\phi_{i,p} = 0.04$}, \emph{$\phi_{i,p} : \phi_{i,c} = 3 : 2$}). The upper row shows various polymer-colloid distributions according to the polymer types and molecular weights (a) PEG \emph{$M_n$} 6,000, (b) PEG \emph{$M_n$} 20,000, (c) PVA \emph{MW} 6,000, (d) PVA \emph{$M_w$} 18,000. The yellow lines represent boundary of stratified layers. If there is no clear boundary, nothing is denoted. The lower rows are estimated relative volume fraction of polymer \emph{$\phi_{p}$} (red circles) and colloid \emph{$\phi_c$} (blue triangles) of two representatives: (e) PEG \emph{$M_n$} 6,000 and (f) PEG \emph{$M_n$} 20,000. The colloidal volume fractions were obtained from SEM images through the ImageJ analysis. The remained volume fraction was considered as polymer volume fraction \emph{$\phi_{p} = 1 - \phi_{c}$}.
}
\end{figure}

\begin{figure}
\includegraphics[scale = 0.7]{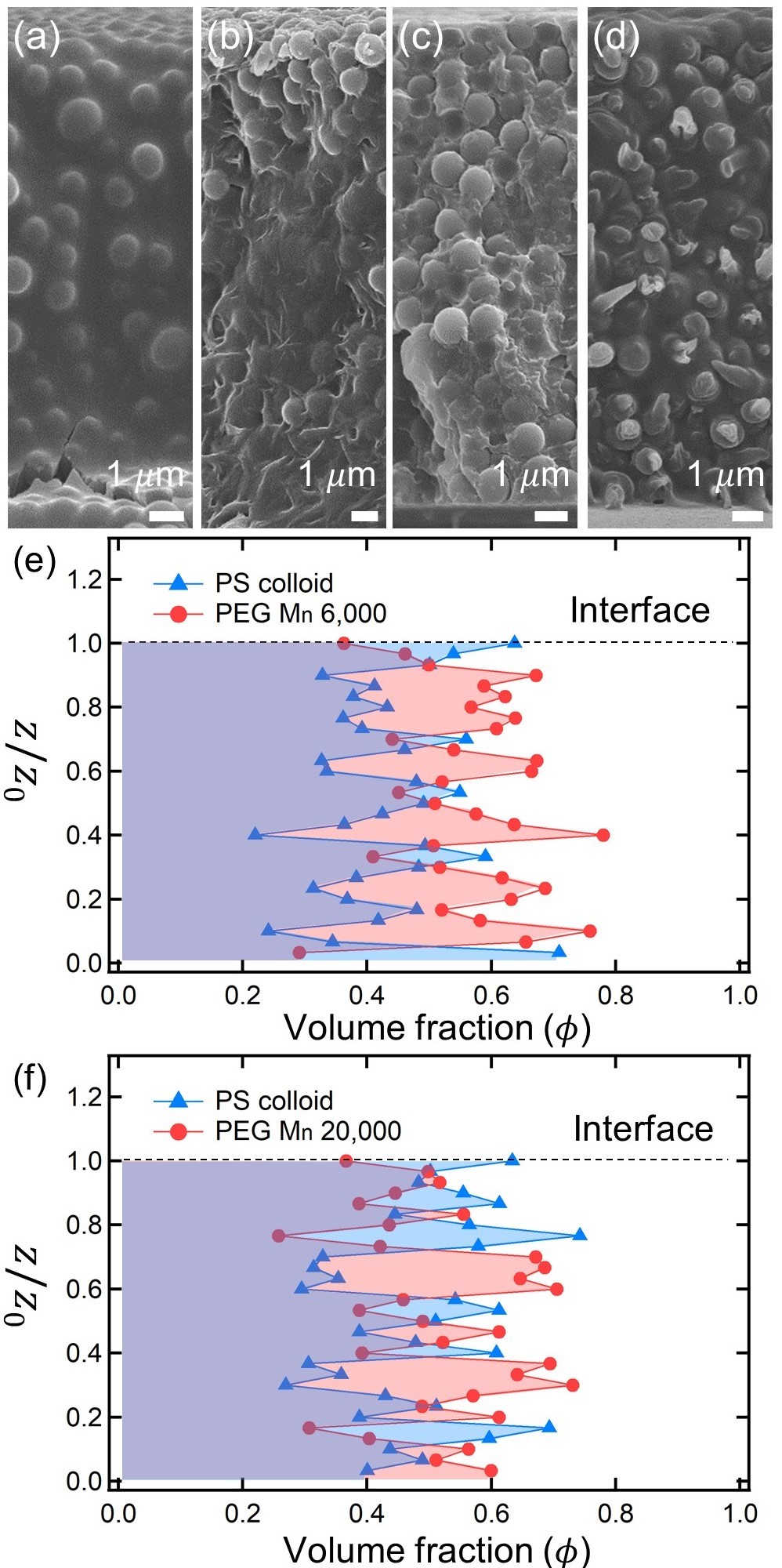}
\caption{\label{fig:epsart}SEM images of dried films formed from polymer-colloid mixtures (\emph{$\phi_{i,p} = 0.01$}, \emph{$\phi_{i,p} : \phi_{i,c} = 3 : 2$}). Distributions of polymer and colloid are shown through the upper row depending on the polymer types and molecular weights (a) PEG \emph{$M_n$} 6,000, (b) PEG \emph{$M_n$} 20,000, (c) PVA \emph{MW} 6,000, (d) PVA \emph{$M_w$} 18,000. There was no clear stratified layer in all four images. The volume fractions of polymer \emph{$\phi_{p}$} (red circles) and colloid \emph{$\phi_{c}$} (blue triangles) of the two dried films were obtained from SEM image analysis: (e) PEG \emph{$M_n$} 6,000 and (f) PEG \emph{$M_n$} 20,000. The volume fractions of colloids are estimated by ImageJ analysis, and the polymer volume fraction was determined by \emph{$\phi_{p} = 1 - \phi_{c}$}.
}
\end{figure}

\subsection{\label{sec:level2}Modified theoretical model of dynamic stratification}
As the solvent evaporated at \emph{Pe} $>$ 1 for both polymer and colloid, the descending air/water interface \emph{$z_{interface}$} compressed the polymer and colloid, and they accumulated near the drying interface. From previous studies [22,34], the transition of the polymer concentration in drying film \emph{$\phi_{p}$}(z,t) can be written as
\begin{eqnarray}
&\phi_{p}(z,t^*) \approx \phi_{i,p}(1+Pe_{p}t^*exp{\bf [}
-\frac{|z-z_{interface}|}{D_{p}/v_{ev}}
{\bf ]}),
\\
&z_{interface}(t^*) = z_{0} - v_{ev}t = (1-t^*)z_{0}
\end{eqnarray}
if \emph{P\'eclet} number of polymer \emph{$Pe_{p}$} $\gg$ 1, where \emph{$t^* = tv_{ev}/z_{0}$} (0 $\leq$ \emph{$t^*$} $\leq$ 1) is the dimensionless time. Here, \emph{$Pe_p$} and diffusion coefficient of polymer \emph{$D_p$} can be expressed as a function of drying time when \emph{$Pe_p$} and \emph{$D_p$} vary slowly. Since the viscosity growth derived from the increased polymer concentration can be accompanied by the kinetic arrest of the colloidal particles, \emph{$t_c^*$} can be determined by the time when the volume fraction of polymer reaches the concentrated regime \emph{$\phi_p = \phi_p^{**}$}. We consider that the colloidal particles at the drying interface (\emph{$z = z_{interface}$}) are kinetically arrested when the polymer fraction reaches \emph{$\phi_p^{**}$} at \emph{$z = z_{interface} - r_{colloid}$}
\begin{eqnarray}
&\phi_{p}(z_{interface} - r_{colloid},t_c^*) = \phi_p^{**}.
\end{eqnarray}

Meanwhile, increasing the concentration gradients of the small polymers can also create the diffusiophoretic drift velocity of larger colloids \emph{$v_{diffusiophoresis}$} [35,36]
\begin{eqnarray}
&v_{diffusiophoresis} = -\frac{9}{4}D_{p}\nabla\phi_{p}
\end{eqnarray}
under the condition of \emph{$R_{colloid}$} $\gg$ \emph{$R_{polymer}$}. From the simple 1D diffusion model, the polymer concentration gradient at the interface is \emph{$\nabla\phi_{p} = -v_{ev} \phi_{interface}/D_{p}$} [37]. This gives the diffusiophoretic velocity of interfacial colloids with the combination of \emph{$\phi_{interface} = \phi_{i,p}(1 + Pe_{p} t^*)$} originating from Eq. (1) at \emph{$z = z_{interface}$},
\begin{eqnarray}
&v_{colloid,interface} \approx \frac{9}{4}v_{ev}\phi_{i,p}(1 + Pe_{p}t^*).
\end{eqnarray}
The time at which the colloid begins to stratify during the evaporation process (\emph{$t_s^*$}) is determined by comparing \emph{$v_{colloid,interface}$} and \emph{$v_{ev}$}. Near the time when evaporation begins, the gradient of polymer concentration is not too large and \emph{$v_{colloid,interface}$} does not overcome \emph{$v_{ev}$}. At this state, both the polymer and colloid simply accumulate at the drying interface. If the concentration gradient of the polymer is large enough for the formation of a higher colloidal diffusiophoretic velocity, however, \emph{$v_{colloid,interface}$} is larger than \emph{$v_{ev}$} and it starts to create stratified layers in the drying film. We consider the time \emph{$t_s^*$} when \emph{$v_{colloid,interface} =  v_{ev}$}, resulting in
\begin{eqnarray}
&v_{colloid,interface}(t_s^*) = v_{ev}.
\end{eqnarray}

The final morphologies of the drying polymer-colloid mixtures are determined by the two competing time scales \emph{$t_s^*$} and \emph{$t_c^*$}. There are three regimes for the predictive analysis of the stratification of polymer-colloid mixtures. The first is \emph{$t_c^*/t_s^*$} $>$ 1, where the downward motion of the colloidal particles appears before \emph{$\phi_{p}(z_{interface}-r_{colloid},t_c^*) = \phi_{p}^{**}$}. The second is \emph{$t_c^*/t_s^*$} $<$ 1, where the polymer volume fraction reaches \emph{$\phi_{p}^{**}$} before the evolution of \emph{$v_{colloid,interface}(t_s^* ) = v_{ev}$}. The third is \emph{$t_s^* \approx 1$}, where \emph{$t_s^*$} reaches to the time at which evaporation ends (\emph{$t^* = 1$}), even though \emph{$t_s^*$} precedes \emph{$t_c^*$}.

\subsection{\label{sec:level2}Comparison of experimental results and theoretical model}
As described above, the prediction for the polymer-colloid stratification can be estimated using the competition between \emph{$t_c^*$} and \emph{$t_s^*$}. For the time dependent volume fraction of the polymer in the drying films, evaporation rates were determined by measuring mass reduction (Fig. SM3). To calculate the time dependent (or concentration dependent) polymer diffusion coefficient, the average volume fractions of polymer in the drying film were used as \emph{$D_{p}$} (See Supplemental Material). The transition volume fraction of semidilute entangled \emph{$\phi_{e}$} to concentrated regime \emph{$\phi^{**}$} in good solvent were determined by the specific viscosity \emph{$\eta_{sp}$} slope transition [27,28,38] in Fig. 3. From the slope transition of semidilute unentangled (\emph{$\eta_{sp}$} $\sim$ \emph{$\phi_{p}^{1.3}$}) to semidilute entangled (\emph{$\eta_{sp}$} $\sim$ \emph{$\phi_{p}^{3.9}$}), \emph{$\phi_{e}$} of the polymer in good solvent was measured. Similarly, \emph{$\phi^{**}$} can be estimated using the slope transition point between the semidilute entangled regime (\emph{$\eta_{sp}$} $\sim$ \emph{$\phi_{p}^{3.9}$}) and the concentrated regime (\emph{$\eta_{sp}$} $\sim$ \emph{$\phi_{p}^{\alpha}$}, where $\alpha$ $>$ 3.9).

\begin{figure}[b]
\includegraphics[scale = 0.9]{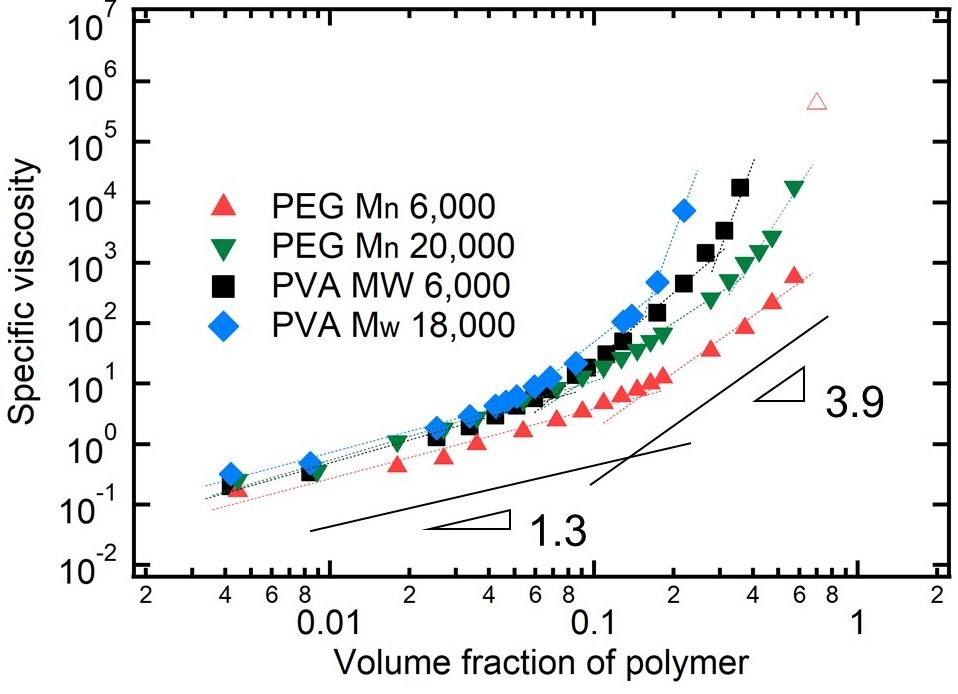}
\caption{\label{fig:epsart}Specific viscosity of four polymer solutions as a function of polymer volume fraction. Polymer volume fraction where it goes to concentrated regime \emph{$\phi^{**}$} is estimated by the slope transition point from 3.9 to larger than 3.9. In case of PEG \emph{$M_n$} 6,000, \emph{$\phi^{**}$} is considered as max solubility ($\approx$ 630 mg/ml at 20$^o$C). As the PEG \emph{$M_n$} 6,000 solution goes to higher than max solubility, it shows abrupt increment of specific viscosity (empty red triangle).
}
\end{figure}

In drying films of polymer-colloid mixtures, the final film morphology can be predicted using the three regimes in the (\emph{$t_s^*$}, \emph{$t_c^*$}) plane. Regime 1 with \emph{$t_c^*$}/\emph{$t_s^*$} $>$ 1 indicates clearly stratified layers in the dried films. Regime 2 represents nonsegregated layers, because \emph{$t_c^*$} appears before \emph{$t_s^*$}. Regime 3 also shows nonstratified layers in the final morphology of the complete dried polymer-colloids mixtures, since \emph{$t_s^*$} appears very close to 1 (\emph{$t_s^*$} $\approx$ 1).

The theoretical predictions based on Eq. (3), Eq. (6) and the experimental stratification results from 8 different systems are presented in Fig. 4. There is quite excellent agreement between the model prediction and experimental results except for the PVA \emph{MW} 6,000 (\emph{$\phi_{i,p} = 0.04$}) system, which also appears to be closest to \emph{$t_c^*/t_s^*$} = 1. This might be due to the air/water interfacial activity of PVA \emph{MW} 6,000 (Fig. SM4), which can make faster \emph{$t_s^*$} under real drying conditions, but it cannot bring \emph{$t_c^*$} forward since \emph{$t_c^*$} is related to the \emph{$z = z_{interface} - r_{colloid}$}, not \emph{$z = z_{interface}$}. To reduce the interfacial activity effect of PVA \emph{MW} 6,000 (\emph{$\phi_{i,p} = 0.04$}) on stratification, we moved the point to deviate from \emph{$t_c^*/t_s^*$} $=$ 1 in our theoretical model by changing \emph{$v_{ev}$}. As it deviates from \emph{$t_c^*/t_s^* = 1$}, the theoretical prediction becomes consistent with the experimental result for PVA \emph{MW} 6,000 (\emph{$\phi_{i,p} = 0.04$}) (Fig. 5).

\begin{figure}
\includegraphics[scale = 0.9]{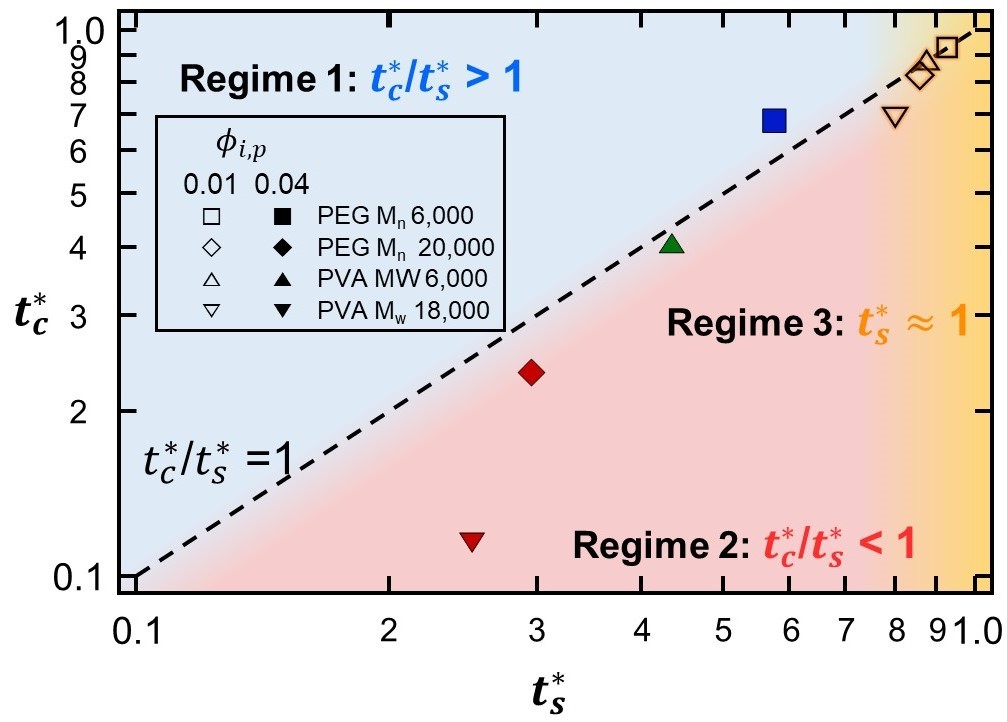}
\caption{\label{fig:epsart}State diagram on the (\emph{$t_s^*$},\emph{$t_c^*$}) plane. The dotted line corresponds to \emph{$t_c^*/t_s^* = 1$}. Theoretical predictions of 8 different systems are denoted as symbols in the diagram, and the experimental results are represented by colors. Blue indicates regime 1 (\emph{$t_c^*/t_s^*$} $>$ 1) where stratified layer expected and red shows regime 2 (\emph{$t_c^*/t_s^*$} $<$ 1). Orange designated regime 3 (\emph{$t_s^* \approx 1$}) (Fig. SM5). The green indicates the intermediate state where stratified layer is observed in experiments while it belongs to regime 2 in model prediction. All data points show overall agreement with one exception, filled green triangle, which also appears close to the \emph{$t_c^*/t_s^* = 1$}.
}
\end{figure}

\begin{figure}
\includegraphics[scale = 0.9]{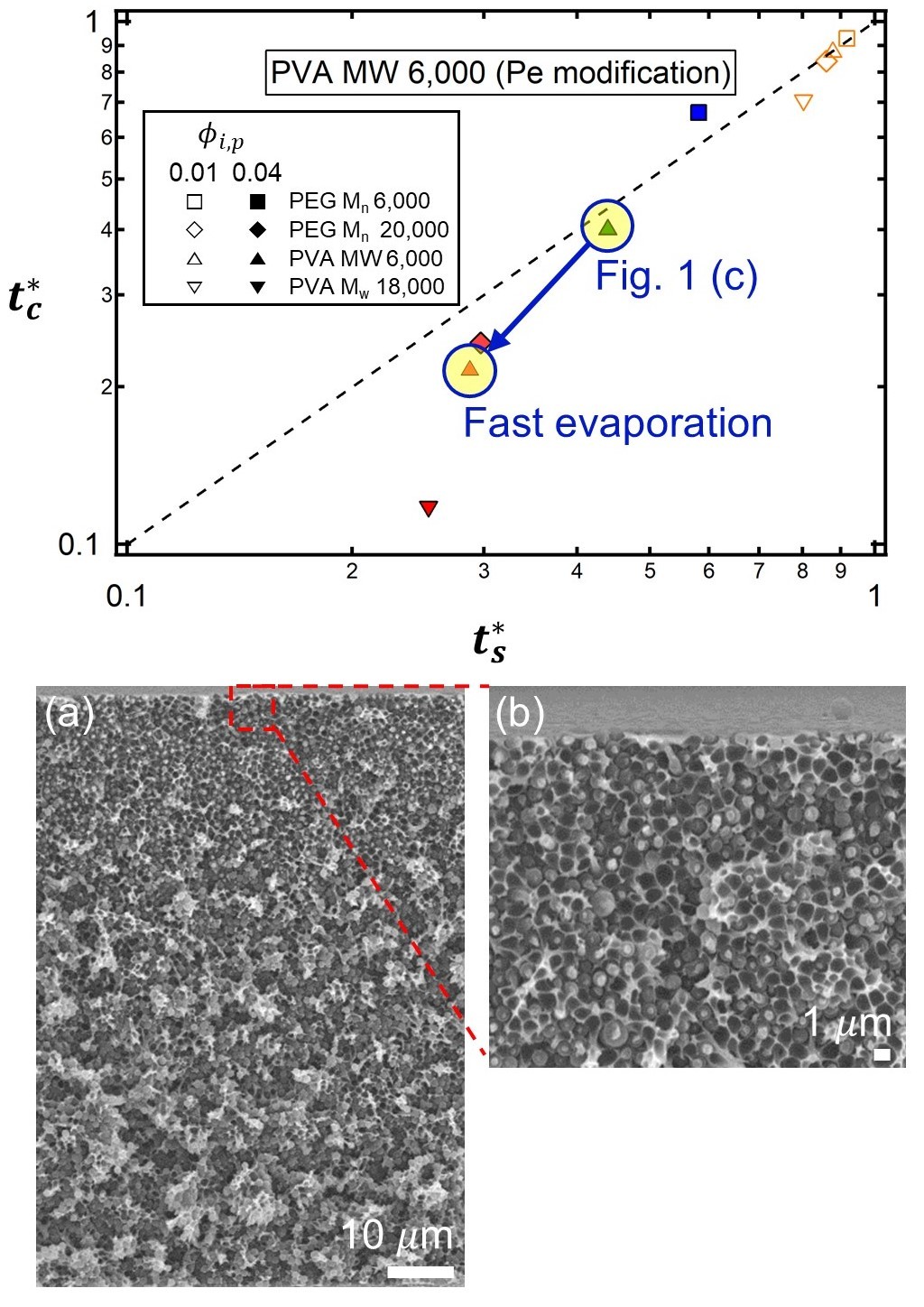}
\caption{\label{fig:epsart}State diagram of PVA \emph{MW} 6,000 (\emph{$\phi_{i,p}$} = 0.04) on the (\emph{$t_s^*$},\emph{$t_c^*$}) plane. The dotted line corresponds to \emph{$t_c^*/t_s^* = 1$}. The filled green triangle deviated from \emph{$t_c^*/t_s^* = 1$} in theoretical model only by increasing \emph{$v_{ev}$}. As it stays away from \emph{$t_c^*/t_s^* = 1$}, the intermediate stratified morphology where stratified layer is observed in experiments while it belongs to regime 2 in model prediction become consistent with model prediction. (a) SEM image of PVA \emph{MW} 6,000 (\emph{$\phi_{i,p}$} = 0.04) at fast evaporation. (b) Top of the cross-sectional SEM image (a). The evaporation rate was controlled by convective flow of air with a relative humidity of 23 \% at ambient temperature.
}
\end{figure}

\subsection{\label{sec:level2}Conditions for polymer-colloid stratification}
To analyze the general conditions for polymer-colloid stratification, we represented \emph{$t_c^*$} and \emph{$t_s^*$} in another experimental parameter. As mentioned above, the polymer-on-top structure can be formed when the two conditions, both \emph{$t_s^*$} $<$ 1 and \emph{$t_c^*/t_s^*$} $>$ 1, are satisfied. From Eq. (3) and Eq. (6), \emph{$t_c^*$} and \emph{$t_s^*$} are (See Supplemental Material)
\begin{eqnarray}
&t_c^* \approx \frac{\frac{\phi^{**}}{\phi_{i,p}}-1}{Pe_{p}(t_c^*)},\\
&t_s^* \approx \frac{\frac{4}{9}\frac{1}{\phi_{i,p}}-1}{Pe_{p}(t_s^*)},
\end{eqnarray}
where \emph{$Pe_{p}(t_c^*)$} and \emph{$Pe_{p}(t_s^*)$} are \emph{Pe} of the polymer at dimensionless time \emph{$t^* = t_c^*$} and \emph{$t^* = t_s^*$} in respectively. The first condition for the stratification to happen, \emph{$t_s^*$} $<$ 1, is
\begin{eqnarray}
&Pe_{p}(t_s^*)\phi_{i,p} > \frac{4}{9} - \phi_{i,p}.
\end{eqnarray}
This follows a condition for similar to that for the inverted stratification of binary colloidal mixtures [21,33].

The second requirement for stratified layers in polymer-colloid mixtures, \emph{$t_c^*/t_s^*$} $>$ 1, can be expressed as
\begin{eqnarray}
&\frac{t_c^*}{t_s^*} \approx \frac{(\frac{\phi^{**}}{\phi_{i,p}}-1)}{(\frac{4}{9}\frac{1}{\phi_{i,p}}-1)} \frac{Pe_{p}(t_s^*)}{Pe_{p}(t_c^*)} > 1  ,\\
&\frac{t_c^*}{t_s^*} \approx \frac{(\frac{\phi^{**}}{\phi_{i,p}}-1)}{(\frac{4}{9}\frac{1}{\phi_{i,p}}-1)} \frac{\eta(t_s^*)}{\eta(t_c^*)} > 1.
\end{eqnarray}
Since \emph{$t_c^*$} and \emph{$t_s^*$} come out when the polymer solution in the semi-dilute entangled regime (close to \emph{$\phi_{p} = \phi^{**}$}), \emph{$\eta(t^*)$} is
\begin{eqnarray}
&\eta(t^*) = (\frac{1-t_e^*}{1-t^*})^{3.9}(\eta_{e}-\eta_{s}) + \eta_{s}
\end{eqnarray}
from Eq. (14) of Supplemental Material, where \emph{$t_e^*$} is the dimensionless time when \emph{$\eta = \eta_{e}$} (viscosity when \emph{$\phi_{p} = \phi_{e}$}) from Eq. (10) of Supplemental Material. If we neglect the last term in Eq. (12),
\begin{eqnarray}
&\frac{t_c^*}{t_s^*} \approx \frac{(\frac{\phi^{**}}{\phi_{i,p}}-1)}{(\frac{4}{9}\frac{1}{\phi_{i,p}}-1)} (\frac{1-t_c^*}{1-t_s^*})^{3.9},\\
&\frac{t_c^*}{t_s^*}(\frac{1-t_s^*}{1-t_c^*})^{3.9} \approx \frac{(\frac{\phi^{**}}{\phi_{i,p}}-1)}{(\frac{4}{9}\frac{1}{\phi_{i,p}}-1)}.
\end{eqnarray}
To satisfy the condition of \emph{$t_c^*/t_s^*$} $>$ 1 for polymer-colloid stratification,
\begin{eqnarray}
&\frac{(\frac{\phi^{**}}{\phi_{i,p}}-1)}{(\frac{4}{9}\frac{1}{\phi_{i,p}}-1)} > 1,\\
&\phi^{**}-\phi_{i,p} > \frac{4}{9} - \phi_{i,p},\\
&\phi^{**} > \frac{4}{9}.
\end{eqnarray}

It is interesting to note that the predicted stratification of the polymer-colloid mixtures does not depend on the drying rate \emph{$v_{ev}$}, or \emph{Pe}, as long as \emph{$Pe \gg 1$}. This tendency also can be seen in Fig. 6, which shows the theoretical predictions of the 8 systems above, with \emph{$v_{ev}$} values changed. Ignoring the data points of  \emph{$Pe_{i,p} \leq 5$}, failing to follow the aforementioned assumption \emph{$Pe \gg 1$}, all the other points belong in same regime once the polymer type and initial volume fraction are determined. This is quite plausible since the increase in polymer concentration near the drying interface accelerates both \emph{$t_c^*$} and \emph{$t_s^*$} in similar order. Thus, it might be hard to create stratified layers in polymer-colloid mixtures only by varying the evaporation rate \emph{$v_{ev}$}, or \emph{Pe}. Altering other properties which can increase \emph{$t_c^*/t_s^*$} larger than 1, such as the interfacial activity of the polymer in Fig. 5 or the gravitational velocity from the density difference in Fig. SM6, could be another solution to achieve stratified layers in polymer-colloid mixtures.

\begin{figure*}
\includegraphics{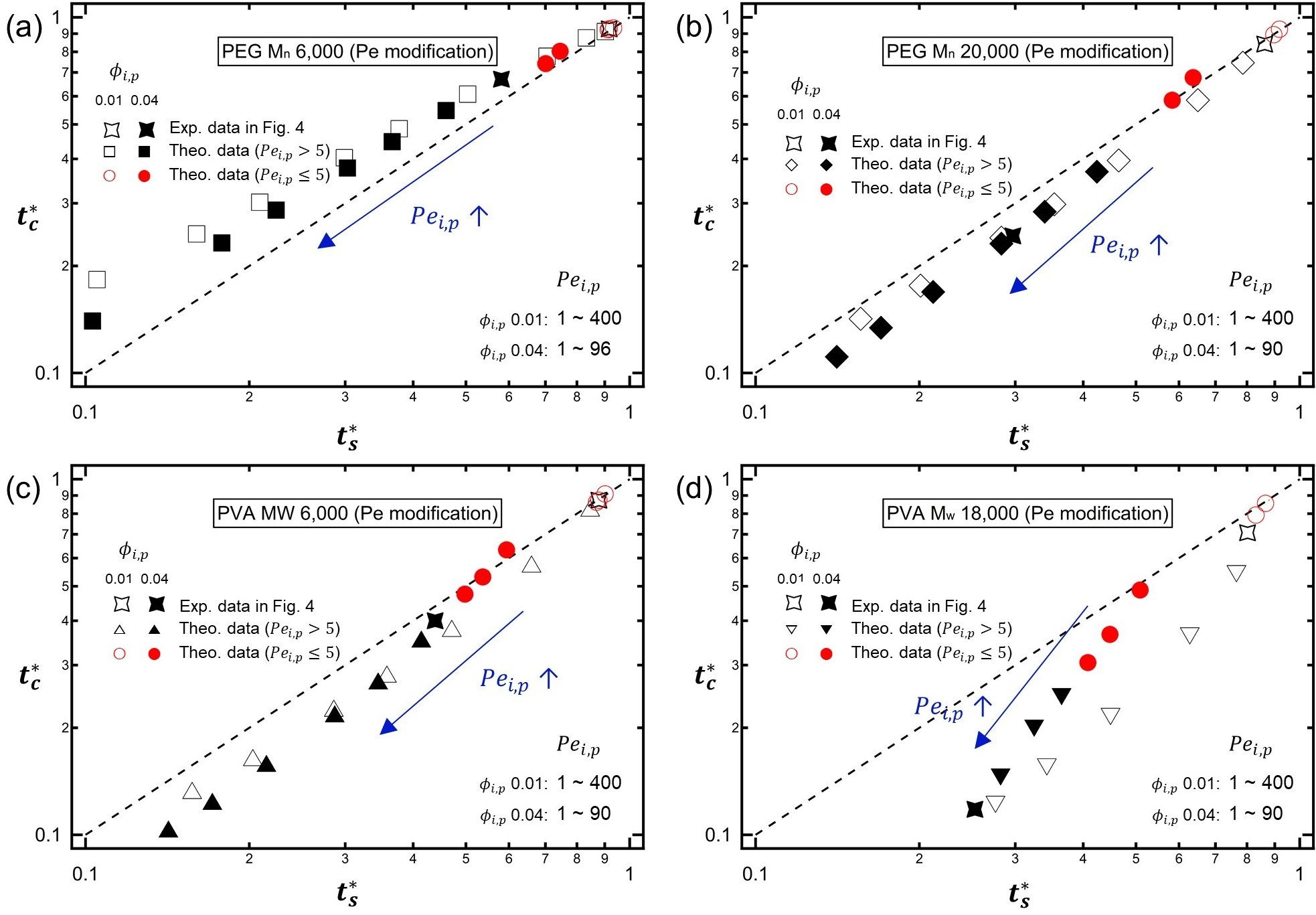}
\caption{\label{fig:epsart}Theoretical prediction of the stratification of 8 different systems on the (\emph{$t_s^*$},\emph{$t_c^*$}) plane with controlled \emph{$v_{ev}$} (or \emph{$Pe_{i,p}$}) (a) PEG \emph{$M_n$} 6,000, (b) PEG \emph{$M_n$} 20,000, (c) PVA \emph{MW} 6,000, (d) PVA \emph{$M_w$} 18,000. As \emph{$Pe_{i,p}$} increases, both \emph{$t_s^*$} and \emph{$t_c^*$} decrease and data points go to left bottom side on the (\emph{$t_s^*$},\emph{$t_c^*$}) plane. Regardless of the polymer type or molecular weight, most of the data points belong in the same regime once the type of polymer and initial volume fraction are determined except the relatively slow drying rate (\emph{$Pe_{i,p} \leq 5$}, red circles).
}
\end{figure*}

\section{Conclusion}
In summary, we demonstrated that dynamic stratification of polymer-colloid mixtures can be achieved by controlling viscosity near the drying interface, which results from increasing polymer concentration. When the polymer-colloid solution evaporates, the polymer starts to increase the solution viscosity near the air/water interface within a relatively very short time, unlike colloidal suspensions. Since the transition in viscosity due to the polymer can cause the kinetic arrest of colloidal particles, which hinders the diffusiophoretic downward motion of colloids, stratified layers are only observed if the formation of a stratified layer precedes the transition in viscosity near the liquid/air interfaces.

Our model calculations for \emph{$t_c^*$} and \emph{$t_s^*$}, inspired by the previous study [22], show that the segregation of polymer-colloid mixtures can only occur under the condition of \emph{$t_c^*/t_s^*$} $>$ 1, unless the solute fraction of the polymer is sufficiently low. The requirement for stratification, \emph{$t_c^*/t_s^*$} $>$ 1, implies that the stratification of polymer-colloid mixtures may not rely on drying rate if \emph{$Pe \gg 1$}, since both \emph{$t_c^*$} and \emph{$t_s^*$} vary in similar order as \emph{$v_{ev}$} changes. Our model calculations are further supported by the consistency between the model prediction and final experimental film morphologies

In more general terms, the consistent results of the experiments and model prediction may shed light on methods of controlling surface enrichment in general solution-cast polymer composites. The ability to predict morphology in a simple nonequilibrium solvent evaporation process is highly desirable for preparing materials whose surface properties are crucial to performance, such as antireflective or organic photovoltaics. Our insights on how polymer concentration affects colloidal dynamics and stratification can be exploited to control segregated layers in solution-cast polymer-colloid mixtures.

\begin{acknowledgments}
This work was supported by the Basic Science Research Program through the National Research Foundation of Korea (Grants NRF-2015R1C1A1A01054180, and NRF-2019R1F1A1059587).
\end{acknowledgments}

\nocite{*}
\bibliography{M_Reference}

\end{document}


\preprint{APS/123-QED}

\title{Stratification of polymer-colloid mixtures via fast nonequilibrium evaporation}


\author{Kyoungmun Lee}
  \author{Siyoung Q. Choi}%
 \affiliation{%
Department of Chemical and Biomolecular Engineering, Korea Advanced Institute of Science and Technology (KAIST), Daejeon 34141, Korea
}%

\maketitle

\renewcommand{\thefigure}{SM\arabic{figure}}

\section*{SURFACE TENSION OF DRYING POLYMER SOLUTION}
To confirm the evolution of polymer concentration gradient (\emph{$Pe_{p} > 1$}) during the fast evaporation process, surface tension of the drying polymer solution (PVA \emph{$M_{w}$}  18,000, \emph{$\phi_{i,p} = 0.04$}) was measured. The evaporation was performed at ambient temperature and a relative humidity of 23\%. As the evaporation progresses, surface tension of the drying solution \emph{$\gamma (t)$} (red circle) became much lower than the saturated surface tension (\emph{$\gamma_{saturated} \approx 60 mN/m$}) in Fig. SM1(a). Thus, the polymers were accumulated near the drying interface in our experimental studiess (\emph{$Pe_{p} > 1$}).

In addition, the homogeneous polymer solution was compared with the drying solution in the pendant drop analysis. Ignoring non-reliable data points of the drying solution (within transparent red square, Bond number $<$ 0.1 [1]) in Fig. SM1(a), \emph{$\gamma (t)$} starts to decrease much rapidly between 30 and 40 minutes. At 30 min after evaporation, the average volume fraction of the polymer \emph{$\phi_{avg}$} reaches 0.06, and it was compared with the homogeneous polymer solution (\emph{$\phi_{i,p} = 0.06$}). As can be seen in Fig. SM1(a), surface tension of the drying solution \emph{$\gamma (t = 30 min)$} (red circle) is lower than surface tension of the homogeneous solution (\emph{$\phi_{i,p} = 0.06$}) (blue triangle). Thus, there are more polymers near the air/water interfaces (appearance of concentration gradient) in the drying solution than the homogeneous solution.

Moreover, the formation of concentration gradient in drying films was further supported by drainage of droplet in Fig. SM1(b) and (c). Although both the drying and the homogeneous droplets have same average polymer volume fraction 0.06, they showed different behavior in drainage. For the drying solution [Fig. SM1(b)], solid skin layer was observed in the drainage while nothing was remained in the homogeneous solution [Fig. SM1(c)].

\begin{figure}
\includegraphics[scale = 0.9]{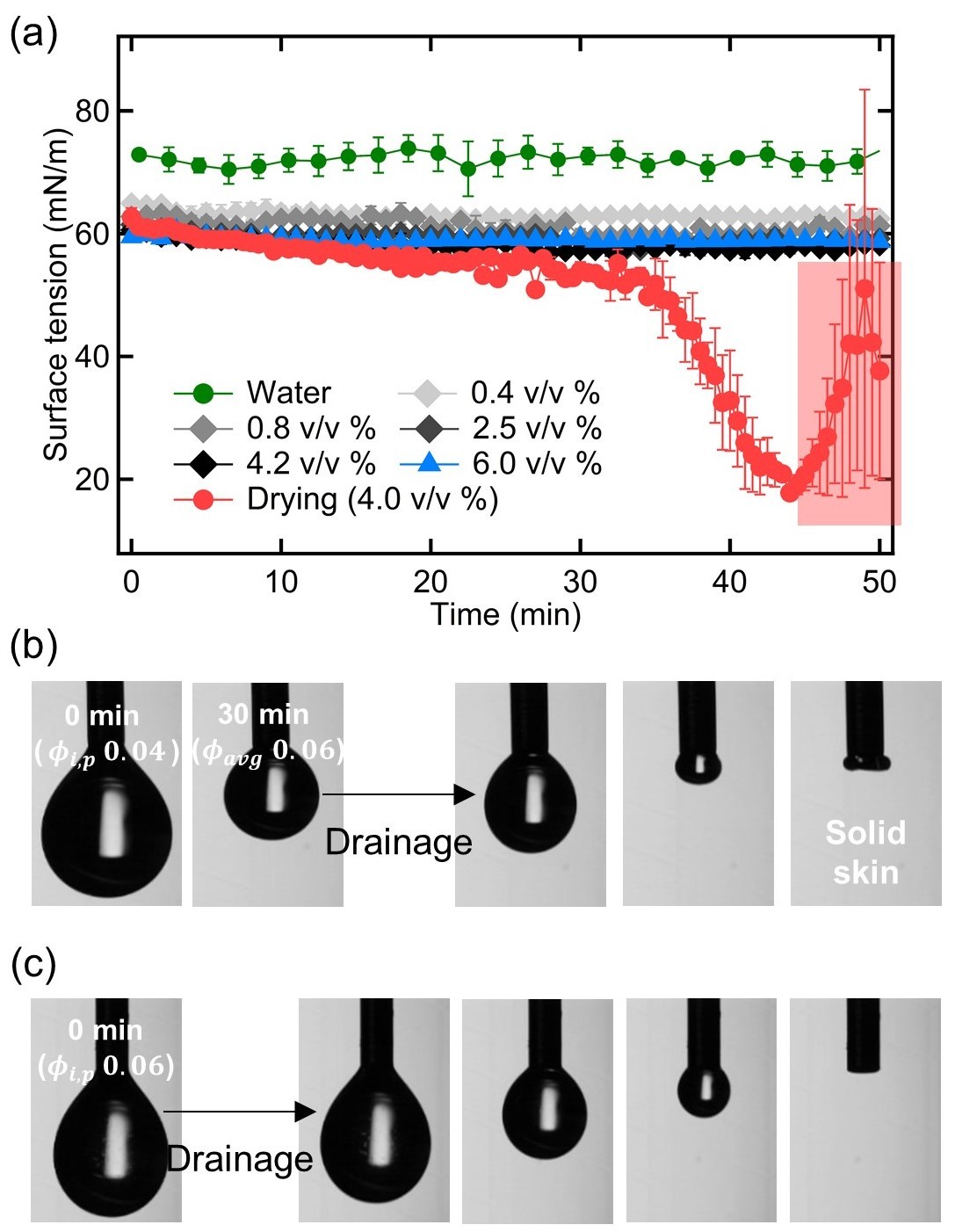}
\caption{\label{fig:epsart}(a) Surface tension of PVA \emph{$M_{w}$}  18,000 solutions. Transition of surface tension of PVA \emph{$M_{w}$}  18,000 (\emph{$\phi_{i,p} = 0.04$}) was investigated to confirm the appearance of polymer concentration gradient near the drying interface. The surface tension of the drying solution became lower than saturated surface tension tension (\emph{$\gamma_{saturated} \approx 60 mN/m$}) as the evaporation progresses. Non-reliable data points are enclosed by red transparent square (Bond number $<$ 0.1). (b) Droplet drainage after 30 min drying (\emph{$\phi_{i,p} = 0.04$}, \emph{$\phi_{avg} = 0.06$}). (c) Droplet drainage of homogeneous solution (\emph{$\phi_{i,p} = 0.06$}).
}
\end{figure}

\section*{RADIUS OF GYRATION OF POLYMER}
Radius of gyration (\emph{$R_g$}) of polymer in good solvent can be estimated by
\begin{eqnarray}
&R_g = \frac{1}{2}bn^{3/5},
\end{eqnarray}
where b is the length of each monomer, and n is the degree of polymerization. From the previous studies [2,3], \emph{$R_g$} of poly(ethylene glycol) and poly(vinyl alcohol) can be calculated by Eq. (1).

\section*{DIFFUSION COEFFICIENT IN DRYING SOLUTIONS}
The concentration dependence (or drying time dependence) of polymer diffusion coefficient \emph{$D_{p}$} was determined using average volume fraction of polymer in drying films. Since the viscosity of polymer-colloid mixtures is similar with the viscosity of pure polymer solutions in Fig. SM2, we only consider polymer concentration effect on the solution viscosity. In the non-draining limit, polymer chains behave as hard spheres and diffusion coefficient \emph{$D_{p} = k_{B}T/6\pi\eta R_{g}$} can be estimated by the variation of viscosity [4]

\begin{align}
&\eta - \eta_{s} \sim \phi_{p}	&&(\text{dilute}, 0 < \phi_{p} < \phi^{*}), \\
&\eta - \eta_{s} \sim \phi_{p}^{1.3} &&(\text{semidilute unentangled}, \phi^{*} < \phi_{p} < \phi_{e}), \\
&\eta - \eta_{s} \sim \phi_{p}^{3.9}	&&(\text{semidilute entangled}, \phi_{e} < \phi_{p} < \phi^{**}), \\
&\eta - \eta_{s} \sim \phi_{p}^{\alpha}	&&(\alpha > 3.9, \text{concentrated}, \phi^{**} < \phi_{p}).
\end{align}
where \emph{$\eta_{s}$} is the solvent viscosity, \emph{$\phi^{*}$} is the overlap volume fraction, \emph{$\phi_{e}$} is the entanglement volume fraction, and \emph{$\phi^{**}$} is the volume fraction of polymer when it reaches concentrated regime. Since initial conditions that we used (\emph{$\phi_{i,p}$} = 0.01, 0.04) is in semidilute regime (Fig. 4), we do not consider the range in dilute regions.

During the evaporation with constant \emph{$v_{ev}$}, average volume fraction of polymer can be written as
\begin{eqnarray}
&\phi_{p,avg} = \frac{V_{i}\phi_{i,p}}{V_{i}-\frac{v_{ev}}{z_{0}}V_{i}t},\\
&\frac{\phi_{p,avg}}{\phi_{i,p}} = \frac{1}{1-t^*},
\end{eqnarray}
where \emph{$V_{i}$} is initial volume of polymer solution and \emph{$t^*$} = \emph{$tv_{ev}/z_{0}$} (\emph{$0 \leq t^* \leq 1$}) is the dimensionless time. Before the polymer solution enters semidilute entangled regime (\emph{$\phi_{i,p} < \phi_{e}$}), the viscosity is
\begin{eqnarray}
&\frac{\eta-\eta_{s}}{\eta_{i}-\eta_{s}} = (\frac{\phi_{p,avg}}{\phi_{i,p}})^{1.3},\\
&\eta = (\frac{\phi_{p,avg}}{\phi_{i,p}})^{1.3}(\eta_{i}-\eta_{s})+\eta_{s},\\
&\eta = (\frac{1}{1-t^*})^{1.3}(\eta_{i}-\eta_{s})+\eta_{s}.
\end{eqnarray}
If the average polymer volume fraction reaches \emph{$\phi_{e}$},
\begin{eqnarray}
&\frac{\phi_{p,avg}}{\phi_{e}} = \frac{\phi_{p,avg}/\phi_{i,p}}{\phi_{e}/\phi_{i,p}} = \frac{1/1-t^*}{1/1-t_e^*} = \frac{1-t_e^*}{1-t^*},
\end{eqnarray}
where \emph{$t_e^*$} is the dimensionless time when \emph{$\eta=\eta_{e}$} (viscosity when \emph{$\phi_{p} = \phi_{e}$}) during drying from Eq. (10). Within the semidilute entangled regime (\emph{$\phi_{e} < \phi_{p,avg} < \phi^{**}$}), \emph{$\eta$} can be determined by
\begin{eqnarray}
&\frac{\eta-\eta_{s}}{\eta_{e}-\eta_{s}} = (\frac{\phi_{p,avg}}{\phi_{e}})^{3.9},\\
&\eta = (\frac{\phi_{p,avg}}{\phi_{e}})^{3.9}(\eta_{e}-\eta_{s})+\eta_{s},\\
&\eta = (\frac{1-t_e^*}{1-t^*})^{3.9}(\eta_{e}-\eta_{s})+\eta_{s}.
\end{eqnarray}
Similarly, \emph{$Pe_{p}$} is determined by the substitution of \emph{$D_{p}$} in \emph{$Pe_{p} = v_{ev} z_{0}/D_{p}$}.

\begin{figure}
\includegraphics[scale = 0.9]{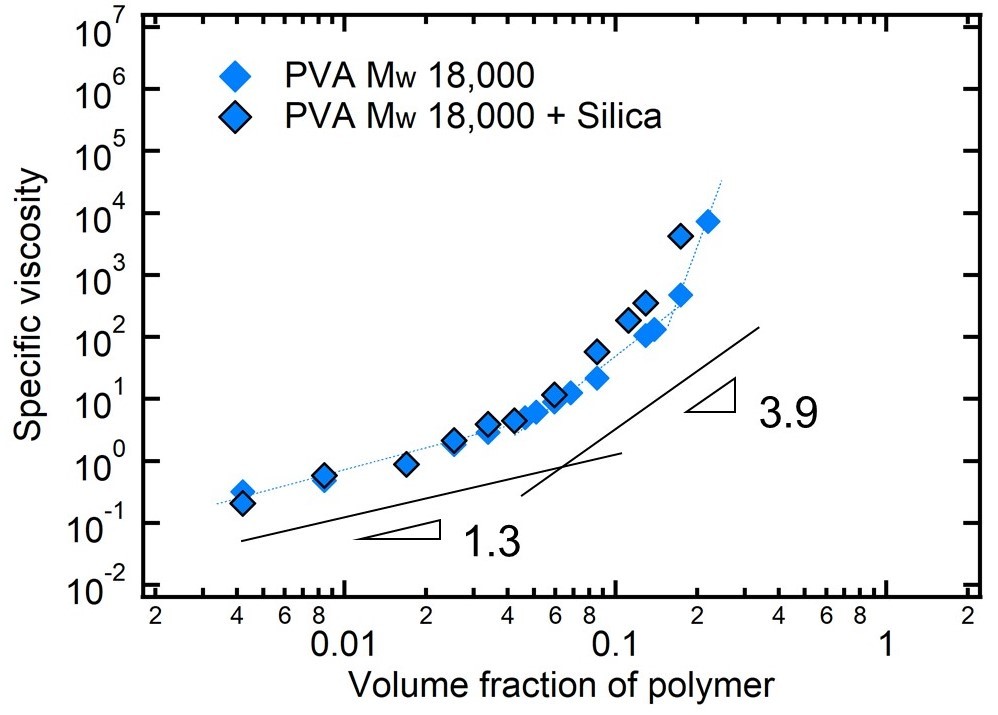}
\caption{\label{fig:epsart}Specific viscosity of polymer-colloid mixture and pure polymer solution. The viscosity of PVA \emph{$M_{w}$}  18,000 + Silica is almost similar with PVA \emph{$M_{w}$}  18,000 before it goes to concentration regime, where the slope changes from 3.9 to larger than 3.9.
}
\end{figure}

\section*{TRANSITION TO ANOTHER EXPERIMENTAL PARAMETER OF $\bm{\emph{$t_c^*$}}$ AND $\bm{\emph{$t_s^*$}}$}
Expressions of \emph{$t_c^*$} and \emph{$t_s^*$} are represented as another experimental parameters. For simplicity of mathematical calculations, we set the dimensionless time \emph{$t_c^*$} as
\begin{eqnarray}
&\phi_{p} (z_{interface},t_c^* ) = \phi_{p}^{**},\\
&\phi_{p} (z_{interface},t_c^*) \approx \phi_{i,p} (1 + Pe_{p}(t_c^*)t_c^*) = \phi_{p}^{**},\\
&t_c^* \approx \frac{\frac{\phi^{**}}{\phi_{i,p}}-1}{Pe_{p}(t_c^*)},
\end{eqnarray}
where \emph{$Pe_{p}(t_c^*)$} is \emph{Pe} of polymer at dimensionless time \emph{$t^* = t_c^*$}.

In similar way, \emph{$t_s^*$} can be written as
\begin{gather}
v_{colloid,interface} (t_s^*) = v_{ev},\\
v_{colloid,interface} (t_s^*) \approx \frac{9}{4} v_{ev} \phi_{i,p} (1 + Pe_{p}(t_s^*) t_s^*) = v_{ev},\\
t_s^* \approx \frac{\frac{4}{9}\frac{1}{\phi_{i,p}}-1}{Pe_{p}(t_s^*)},
\end{gather}
where \emph{$Pe_{p}(t_s^*)$} is \emph{Pe} of polymer at dimensionless time \emph{$t^* = t_s^*$}.

\begin{figure}
 \textbf{EVAPORATION RATE OF POLYMER-COLLOID MIXTURES}\par\medskip
\includegraphics[scale = 0.9]{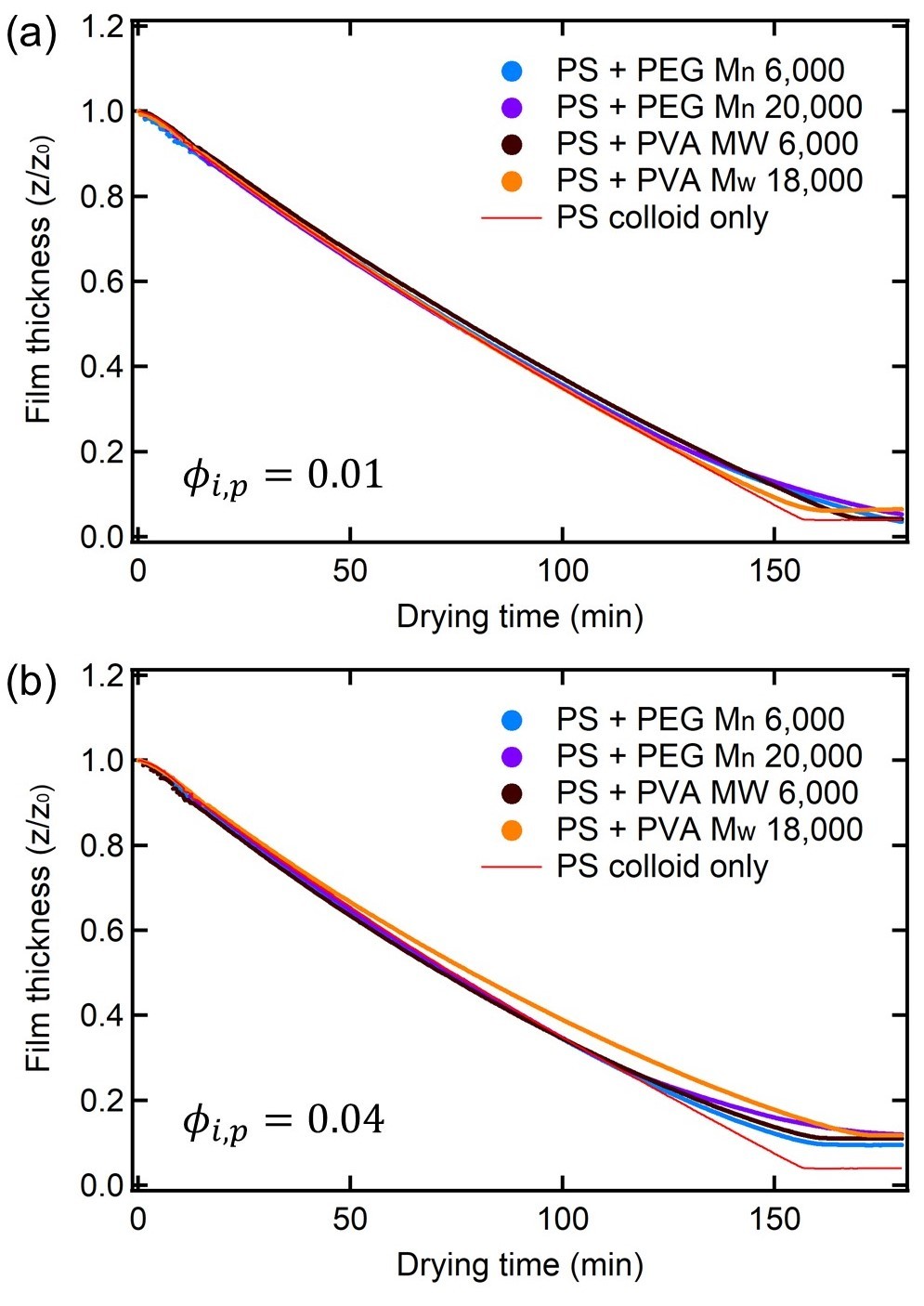}
\caption{\label{fig:epsart}Normalized film thickness of drying polymer-colloid mixtures. They show almost constant evaporation velocity \emph{$v_{ev}$} except near the end of evaporation time. (a) \emph{$\phi_{i,p} = 0.01$} (b) \emph{$\phi_{i,p} = 0.04$}. The evaporation rate \emph{$v_{ev}$} was determined by the slope of the measured film thickness during 10 minutes of initial drying time.
}
\end{figure}

\begin{figure}
 \textbf{AIR/WATER INTERFACIAL ACTIVITY OF PVA $\bm{\emph{$MW$}}$ 6,000}\par\medskip
\includegraphics[scale = 0.9]{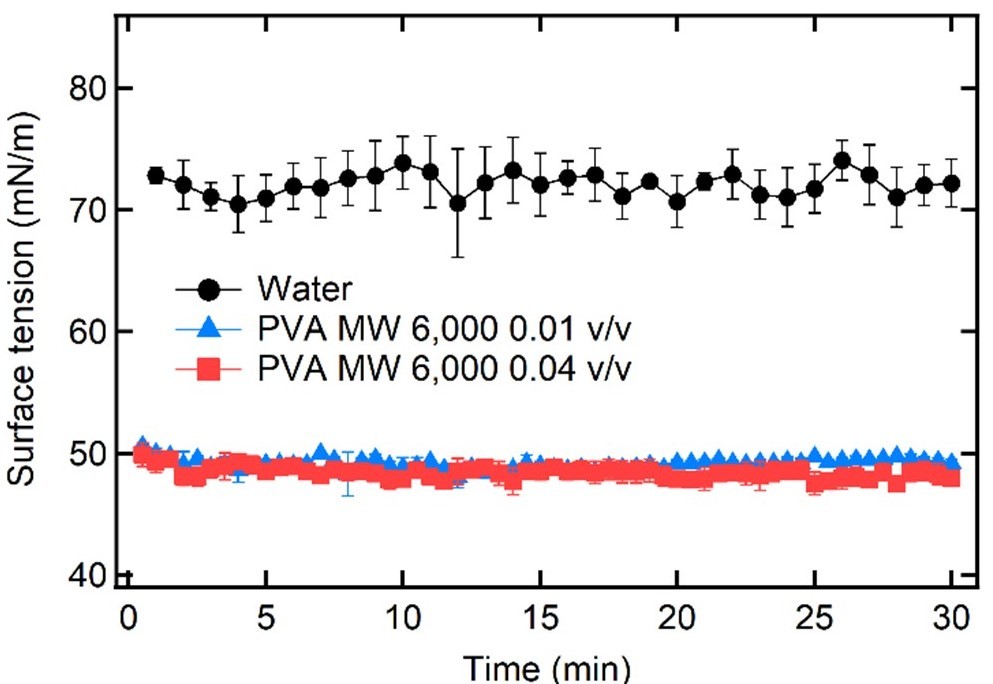}
\caption{\label{fig:epsart}Surface tension of PVA \emph{$MW$} 6,000 solution. Both \emph{$\phi_{p} =$} 0.01 and 0.04 shows similar surface tension value. This means that interfacial activity of PVA \emph{$MW$} 6,000 was almost saturated before the \emph{$\phi_{p}$} reached 0.01. In addition, PVA \emph{$MW$} 6,000 goes to air/water interface within a very short time ($<$ 3 min).
}
\end{figure}

\section*{CRITERIA FOR REGIME 3 ($\bm{\emph{$t_s^* \approx 1$}}$)}
To determine whether the system is in regime 3 (how close should \emph{$t_s^*$} be to 1), the length of initial accumulation zone \emph{$D_{i,p}/v_{ev}$}, where \emph{$D_{i,p}$} is the initial diffusion coefficient of polymer, was compared with remained film height \emph{$(1-t^*) z_0$} at \emph{$t^* = t_s^*$}. If the initial accumulation zone already reaches bottom substrate at \emph{$t^* = t_s^*$}, we consider that the system belongs to regime 3. For regime 3, 
\begin{eqnarray}
&(1-t_s^*) z_0 < D_{i,p}/v_{ev},\\
&t_s^* > 1 - 1/Pe_{i,p}.
\end{eqnarray}
Among the 8 systems, 4 systems having \emph{$\phi_{i,p} = 0.01$} are close to regime 3 (Fig. SM5).
\begin{figure}[t]
\includegraphics[scale = 0.9]{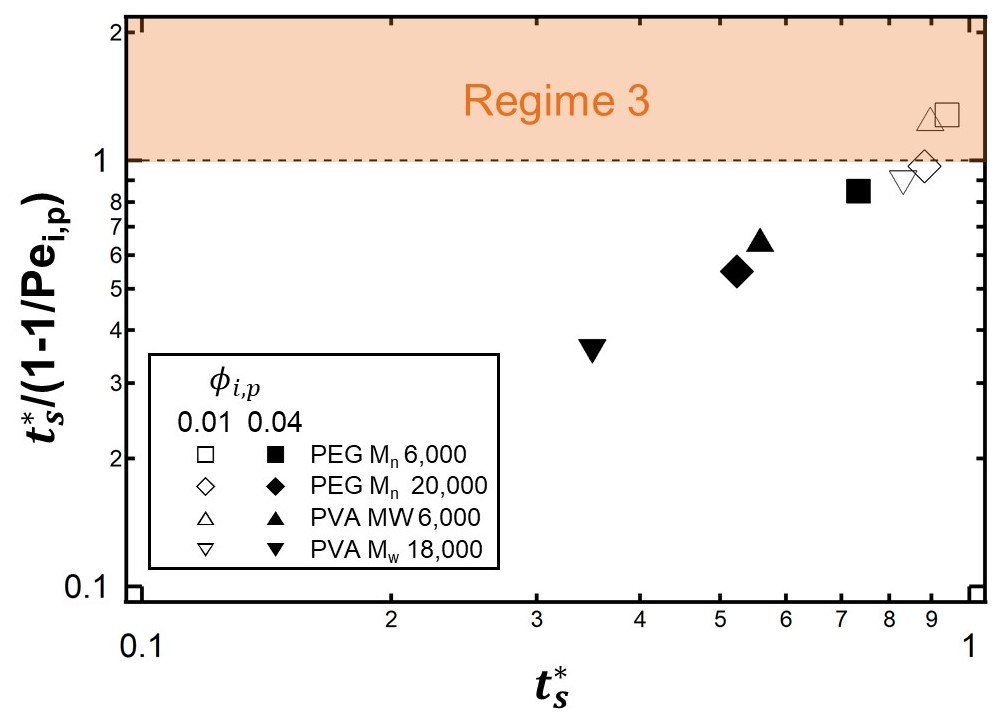}
\caption{\label{fig:epsart}Comparison of the initial accumulation zone and remained film height at \emph{$t^* = t_s^*$}. Systems having \emph{$\phi_{i,p} = 0.01$} show very close to regime 3, while \emph{$t_s^*$} of \emph{$\phi_{i,p}$} = 0.04 systems quite far from \emph{$1 - 1/Pe_{i,p}$}. 
}
\end{figure}

\section*{GRAVITY EFFECT ON STRATIFICATION}
For the formation of stratified layers in dried films which cannot achieve stratification only by changing \emph{$v_{ev}$}, gravitation effect was investigated. Theoretical calculation of \emph{$t_s^*$} was performed with the assumption that medium density \emph{$\rho_{0}$} is almost constant during drying (\emph{$\rho_{0} \approx 1 g/ml$}). The transition of medium viscosity \emph{$\eta$} was only considered for the additional gravitational velocity \emph{$v_{g}$}. Gravitational velocity \emph{$v_{g}$} is
\begin{eqnarray}
&v_{g} = - \frac{2r^2 (\rho-\rho_{0})g}{9\eta},
\end{eqnarray}
where \emph{r} is radius of colloid, \emph{$\rho$} is colloidal density and \emph{g} is gravitational acceleration. Thus, the velocity of interfacial colloids with the addition of gravitational effect can be written as
\begin{align}
&v_{grav-colloid,interface} \approx \frac{9}{4} v_{ev} \phi_{i,p} (1 + Pe_{p} t^*) + v_{g}.
\end{align}
We determine that \emph{$t_s^*$} emerges when \emph{$v_{grav-colloid,interface} = v_{ev}$} and \emph{$t_c^*$} equals the value without considering gravitational velocity.

In case of PS colloid that was used in Fig. 1 and Fig. 2, \emph{$v_{g}$} can be neglected since the density of PS (\emph{$\rho_{PS} = 1.05 g/ml$}) is quite similar to the water density (\emph{$\rho_{water} = 1.00 g/ml$}). If the type of colloid is changed to silica (\emph{$\rho_{silica} = 2 g/ml$}) with PVA \emph{$M_w$} 18,000 (\emph{$\phi_{i,p} = 0.04$}), \emph{$t_c^*/t_s^*$} goes next to 1 and this can change final dried film morphology from nonstratified layer to clearly stratified layer in Fig. SM5.

\begin{figure}
\includegraphics[scale = 0.9]{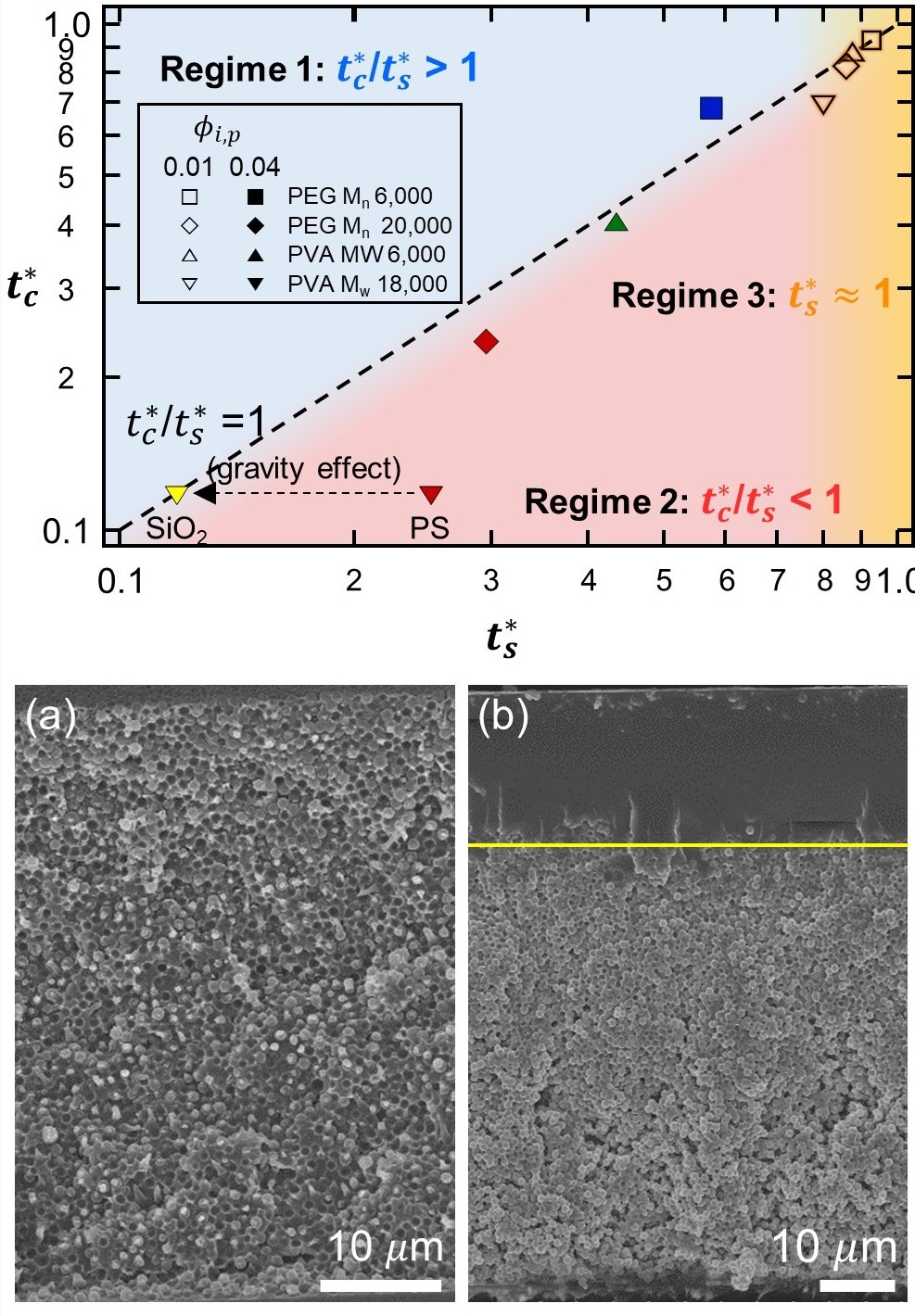}
\caption{\label{fig:epsart}Gravity effect on stratification of polymer-colloid mixtures (PVA \emph{$M_w$} 18,000, \emph{$\phi_{i,p} = 0.04$}, \emph{$\phi_{p} : \phi_{c} = 3 : 2$}). All the experimental conditions are same except for the colloidal type, (a) PS (\emph{$\rho_{PS} = 1.05 g/ml$}) and (b) silica (\emph{$\rho_{silica} = 2.00 g/ml$}). For the theoretical calculation, the gravitational velocity from density difference is added to \emph{$v_{colloid,interface}$}. SEM image of (a) is same with Fig. 2(d). Clear stratification can be observed in colloidal silica and polymer mixture, while randomly distributed structure was obtained in colloidal PS and polymer mixture.
}
\end{figure}

\section*{DETAILS OF THE EXPERIMENTS}
\subsection{\label{sec:level2}Materials}
Glass slide substrates of 27 mm $\times$ 76 mm in size were purchased from Duran. Ultrapure water was produced by Millipore ICW-3000 water purification system ($>$ 18 M$\Omega$). Polystyrene microsphere with a diameter of 1.0 $\mu$m ($\pm$ 0.04 $\mu$m-s.d., 10.14$\%$ solid content in water) was provided by Bang’s laboratories, Inc. and 1.0 $\mu$m silica particles ($\pm$ 0.05 $\mu$m-s.d., 10$\%$ solid content in water) was purchased from Polysciences, Inc. Both PS and silica colloidal particles purified at least three times with deionized water. It was first centrifuged, and then supernatant was replaced by purified water. Poly(ethylene glycol) (PEG) \emph{$M_n$} (number average molecular weight) 6,000 gmol$^{-1}$, PEG \emph{$M_n$} 20,000 gmol$^{-1}$, and poly(vinyl alcohol) (PVA) \emph{$M_w$} (weight average molecular weight) 13,000-23,000 gmol$^{-1}$ (degree of hydrolysis 98$\%$) were purchased from Sigma-Aldrich. PVA \emph{$MW$} 6,000 gmol$^{-1}$ (degree of hydrolysis 80$\%$) was purchased from Polysciences, Inc. The desired concentration of PVA solution was prepared by heating the PVA-water mixtures at 85$^o$C for 1 h, then cooling the system to ambient temperature. Highly concentrated PVA solution was heated at 85$^o$C for 12 h to obtain homogeneous solutions.

\hfill

\subsection{\label{sec:level2}Dried film formation}
Glass substrates (10 mm $\times$ 10 mm) were cleaned in acetone in a sonicating bath for 10 min. They were dried using nitrogen gas and placed in a drying oven (Changshin Science) for 3 min to completely remove acetone. To achieve the required volume ratio (\emph{$\phi_{p} : \phi_{c} = 3 : 2$}), polymers and colloids were mixed to prepare polymer-colloid mixtures, which was then diluted by the addition of deionized water until the desired initial concentration. Solution mixtures of polymer-colloid were deposited on the glass substrates with an initial wet thickness of approximately 1.25 mm. The samples were dried at ambient temperature with a relative humidity of 23$\%$ (measured by TFA digital thermos-hygrometer, Daedeok scientific), leading to the \emph{$Pe > 1$} for both the colloid and polymer. Relative humidity was achieved by placing a bottle filled with water or purging with N$_{2}$ gas in the drying chamber.

\subsection{\label{sec:level2}Structure characterization}
The final morphology of dried films was characterized by observing the cross-sectional images through the scanning electronic microscopy (SEM) (Hitachi, Japan). The samples were coated with Osmium coater (HPC-1SW, Vacuum Device) before taking images to prevent charging effect. In addition, imaging processing of SEM images in ImageJ software further analyzed the spatial distribution of polymers and colloids.

\subsection{\label{sec:level2}Dynamic viscosity of polymer solutions}
After preparing polymer solutions of various concentrations (two molecular weights of both PEG and PVA, from 0.5 to 30 - 60 v/v$\%$ depending on the system), rheological properties were identified using a conventional rheometer (MCR 302, Anton Paar, Austria) with cone plate geometry (0.1 mm gap, 50 mm diameter). The viscoelastic properties of each solution were characterized by dynamic viscosity measurements within the shear rate range from 0.01 s$^{-1}$ to 1000 s$^{-1}$. Most dispersions exhibited Newtonian fluid behavior within the measured shear rate and the viscosity value at the lowest shear rate was considered zero shear viscosity.

\nocite{*}
\bibliography{S_Reference}